\documentclass[a4paper]{article}

\usepackage[ninepoint]{itgspeech2023}    
\usepackage{times}            
\usepackage[english]{babel}   
\usepackage[utf8]{inputenc}
\usepackage[T1]{fontenc}      
\usepackage[sort&compress,numbers]{natbib}	
\usepackage{amsmath,amssymb}
\usepackage{amsfonts,mathtools}
\DeclareMathAlphabet{\mathcal}{OMS}{cmsy}{m}{n}
\usepackage{bm}
\usepackage{graphicx}
\usepackage{adjustbox}
\usepackage[colorlinks=false,pdfborder={0 0 0}]{hyperref}
\usepackage{doi}
\usepackage{siunitx}
\usepackage{booktabs}
\usepackage{array}
\usepackage{tabularx}
\usepackage[svgnames,table]{xcolor}

\usepackage{blindtext}
\usepackage{balance}
\usepackage{microtype}

\newcommand\blfootnote[1]{%
  \begingroup
  \renewcommand\thefootnote{}\footnote{#1}%
  \addtocounter{footnote}{-1}%
  \endgroup
}

\usepackage{pgfplots}
\usepackage{tikz}
\pgfplotsset{compat=newest}
\pgfplotsset{plot coordinates/math parser=false} 
\usepgfplotslibrary{patchplots}
\usepgfplotslibrary{colormaps}%

\DeclareMathSymbol{\shortminus}{\mathbin}{AMSa}{"39}
\newcommand{\y}{\mathbf{y}}
\newcommand{\x}{\mathbf{x}}
\newcommand{\n}{\mathbf{n}}
\newcommand{\R}{\mathbf{R}}
\newcommand{\Ry}{\R_y}
\newcommand{\Ryest}{\hat{\R}_y}

\newcommand{\Rx}{\R_x}

\newcommand{\Rn}{\R_n}
\newcommand{\Rnest}{\hat{\R}_n}

\newcommand{\evec}{\mathbf{e}}
\newcommand{\h}{\mathbf{h}}
\newcommand{\hest}{\hat{\mathbf{h}}}
\newcommand{\inv}{^{-1}}
\newcommand{\w}{\mathbf{w}}

\newcommand{\refsub}{\mathrm{ref}}

\newcommand{\ones}[2]{\mathbf{1}_{#1\times#2}}
\newcommand{\SC}{{\rm SC}}
\newcommand{\mSNR}{{\rm mSNR}}
\newcommand{\esub}{{\rm E}}
\newcommand{\hsub}{{\rm H}}
\newcommand{\Hest}{\hat{\mathbf{H}}}
\newcommand{\avec}{\boldsymbol{\alpha}}
\newcommand{\Amat}{\mathbf{A}}
\newcommand{\Bmat}{\mathbf{B}}

\newcommand{\Tsix}{\mathrm{T}_{\mathrm{60}}}

\title{BRUDEX Database: \textit{B}inaural \textit{R}oom Impulse Responses with \textit{U}niformly \textit{D}istributed \textit{Ex}ternal Microphones}
\author{Daniel Fejgin$^*$, Wiebke Middelberg$^*$, Simon Doclo}

\address{University of Oldenburg, Department of Medical Physics and Acoustics and Cluster of Excellence Hearing4all, Oldenburg, Germany\\
Email: \{\href{mailto:daniel.fejgin@uni-oldenburg.de}{\nolinkurl{daniel.fejgin}}, \href{mailto:wiebke.middelberg@uni-oldenburg.de}{\nolinkurl{wiebke.middelberg}}, \texttt{simon.doclo\}@uni-oldenburg.de}\\
Web: \href{https://uol.de/en/mediphysics-acoustics/sigproc}{\nolinkurl{www.sigproc.uni-oldenburg.de}}}

\begin{document}

\maketitle

\begin{abstract} 
There is an emerging need for comparable data for multi-micro\-phone processing, particularly in acoustic sensor networks. However, commonly available databases are often limited in the spatial diversity of the microphones or only allow for particular signal processing tasks. In this paper, we present a database of acoustic impulse responses and recordings for a binaural hearing aid setup, 36 spatially distributed microphones spanning a uniform grid of \qtyproduct[product-units = bracket-power]{5 x 5}{\metre^{2}} and 12 source positions. This database can be used for a variety of signal processing tasks, such as (multi-microphone) noise reduction, source localization, and dereverberation, as the measurements were performed using the same setup for three different reverberation conditions ($\mathrm{T}_{\mathrm{60}}\approx \{310, 510, 1300\}$ \unit{\milli\second}). The usability of the database is demonstrated for a noise reduction task using a minimum variance distortionless response beamformer based on relative transfer functions, exploiting the availability of spatially distributed microphones.

\blfootnote{$^*$Equal contribution. This work was funded by the Deutsche Forschungsgemeinschaft (DFG, German Research Foundation) - Project ID 352015383 (SFB 1330 B2) and Project ID 390895286 (EXC 2177/1).}
\end{abstract}


\section{Introduction}\label{sec:intro}

Signal processing for acoustic sensor networks is a field of in\-creasing interest \cite{MarkovichGolan2015ASN,Tavakoli2016ASN,Cobos2017survey,Koutrouvelis2018WASN,Zhang2019RTF,GoesslingWASPAA2019} as spatially distributed microphones allow for a more diverse spatial sampling of the sound field than compact microphone arrays. To enable reproducible and comparable research, publicly available databases that allow for different signal processing tasks, such as speech enhancement or sound source localization, are of utmost importance. Despite the availability of simulation tools for the generation of room impulse responses (RIRs), e.g., \cite{Allen1979,SMIR}, measured RIRs and real-world recordings are indispensable to evaluate the performance of algorithms.
Nowadays, there is a variety of multi-microphone RIR databases available, e.g., considering linear arrays \cite{Hadad2014database}, widely distributed microphones \cite{Stewart2010database,Koyama2021meshrir}, head-mounted devices like hearing aids \cite{Kayser2009,Denk2018database}, and head-mounted or body-worn microphones together with distributed microphones \cite{Woods2015database,Corey2019database,dietzen2023myriad}.

\begin{figure}[t]
	\centerline{\includegraphics[width=0.98\columnwidth,trim={0.1cm 0.15cm 0cm 0.3cm},clip]{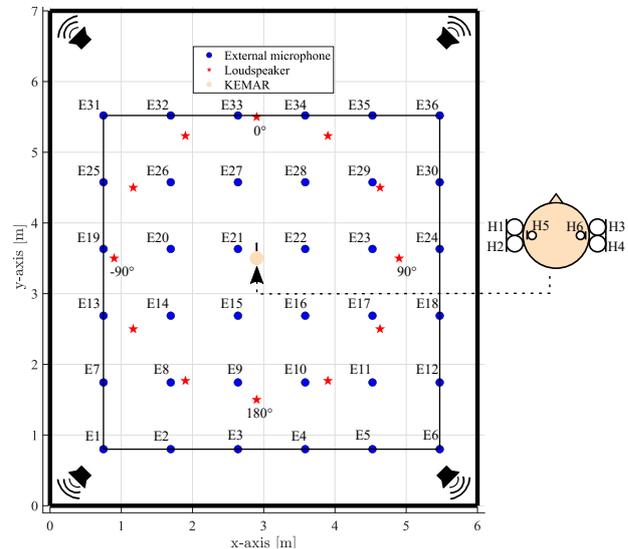}}
    \vskip-0.15cm
	\caption{Placement of external microphones (\raisebox{-1pt}{\scalebox{1.6}{\textcolor{blue}{$\bullet$}}}), loudspeakers (\raisebox{-1pt}{\scalebox{1.6}{\textcolor{red}{$\star$}}}) and KEMAR dummy head with four hearing aid microphones and two in-ear microphones (see close-up) in a laboratory with variable reverberation time.}
	\label{fig:Setup}
    \vskip-0.2cm%
\end{figure}

In this paper, we present a new, complementary database, referred to as {\bf B}inaural {\bf R}oom Impulse Responses with {\bf U}niformly {\bf D}istributed {\bf Ex}ternal microphones (BRUDEX). It consists of a total of about 1500 measured RIRs for three different reverberation conditions, 12 source positions, binaural hearing aids on a dummy head (four hearing aid microphones and two in-ear microphones), and spatially distributed microphones at 36 uniformly distributed positions (see Figure \ref{fig:Setup}). To distinguish between the head-mounted microphones (i.e., hearing aid and in-ear microphones) and the spatially distributed microphones, we will refer to the latter as external microphones (eMics). The database was recorded in an acoustic laboratory at the University of Oldenburg (see Figure \ref{fig:Photo}). The reverberation condition in the laboratory can be set by means of curtains and absorber panels, which are mounted to the walls and the ceiling. The RIRs were measured for three reverberation conditions, keeping the microphone and loudspeaker configuration unchanged while varying the reverberation condition. Besides measured RIRs, the database contains re\-cordings of speech and noise signals for all mentioned conditions. A summary of the content of the database is presented in Table \ref{tab:summaryDatabase}.  
\begin{table*}[t]
\caption{Content of the BRUDEX database.}\label{tab:summaryDatabase}\vskip0.05cm
\begin{tabular}{@{}l >{\raggedright\arraybackslash}m{3.5cm} >{\centering}m{5.75cm} >{\centering\arraybackslash}m{4cm}@{}}\toprule[0.05cm]
~ & RIR & speech signal & noise signal\\\midrule[0.05cm]
Reverb condition & low, medium, high & low, medium, high & low, medium, high\\[0.25cm]
Type & -- & \begin{tabular}{>{\centering}m{1cm} >{\centering}m{1cm} >{\centering}m{1cm} >{\centering}m{1cm}}female&female&male&male\end{tabular} & \begin{tabular}{ccc}babble&cafeteria&white\end{tabular}\\[0.25cm]
Duration [s] & 2 & \begin{tabular}{>{\centering}m{1cm} >{\centering}m{1cm} >{\centering}m{1cm} >{\centering}m{1cm}}20&30&20&30\end{tabular} & 20 \\[0.25cm]
DOA [deg] & -180:30:150 & -180:30:150 & quasi-diffuse \\\bottomrule[0.05cm]
\end{tabular}
\end{table*}
The BRUDEX database allows for a variety of signal processing tasks using spatially distributed microphones: First, the database obviously allows for multi-microphone noise reduction and speech enhancement, as both measured RIRs as well as separate speech and noise recordings are available. Second, the database allows for dereverberation in different acoustic conditions, as measurements for different reverberation conditions are available. Third, the database allows for source localization, as all microphone and loudspeaker positions are calibrated. 

In Sections \ref{sec:setup} and \ref{sec:recording__signalAcquisition}, we provide details on the setup, the measurement conditions and the calibration and measurement procedures. In Section 4, we provide practical information about the accessibility. To demonstrate the usability of the BRUDEX database and the compatibility with the database in \cite{Kayser2009}, in Section \ref{sec:application} we construct an acoustic scenario consisting of a target speech source and diffuse-like background noise using recordings from the four hearing aid microphones and three external microphones. We consider a binaural minimum variance distortionless response beamformer steered by the relative transfer function vector \cite{Doclo2018}, which is either estimated blindly from the microphone signals or obtained from database RIRs. The simulation results show a good accordance with the results from recent literature. 

\section{Setup and Spatial Configurations}
\label{sec:setup}%

In this section, we provide an overview of the acoustic setup and the calibration procedure for the BRUDEX database. In Section \ref{subsec:room}, we describe the laboratory and its acoustic properties and in Section \ref{subsec:recording__geometry}, we describe the configuration of the loudspeakers and microphones.

\subsection{Acoustic Properties of the Laboratory}\label{subsec:room}
\begin{figure}
    \centerline{
    \includegraphics[width=0.95\linewidth]{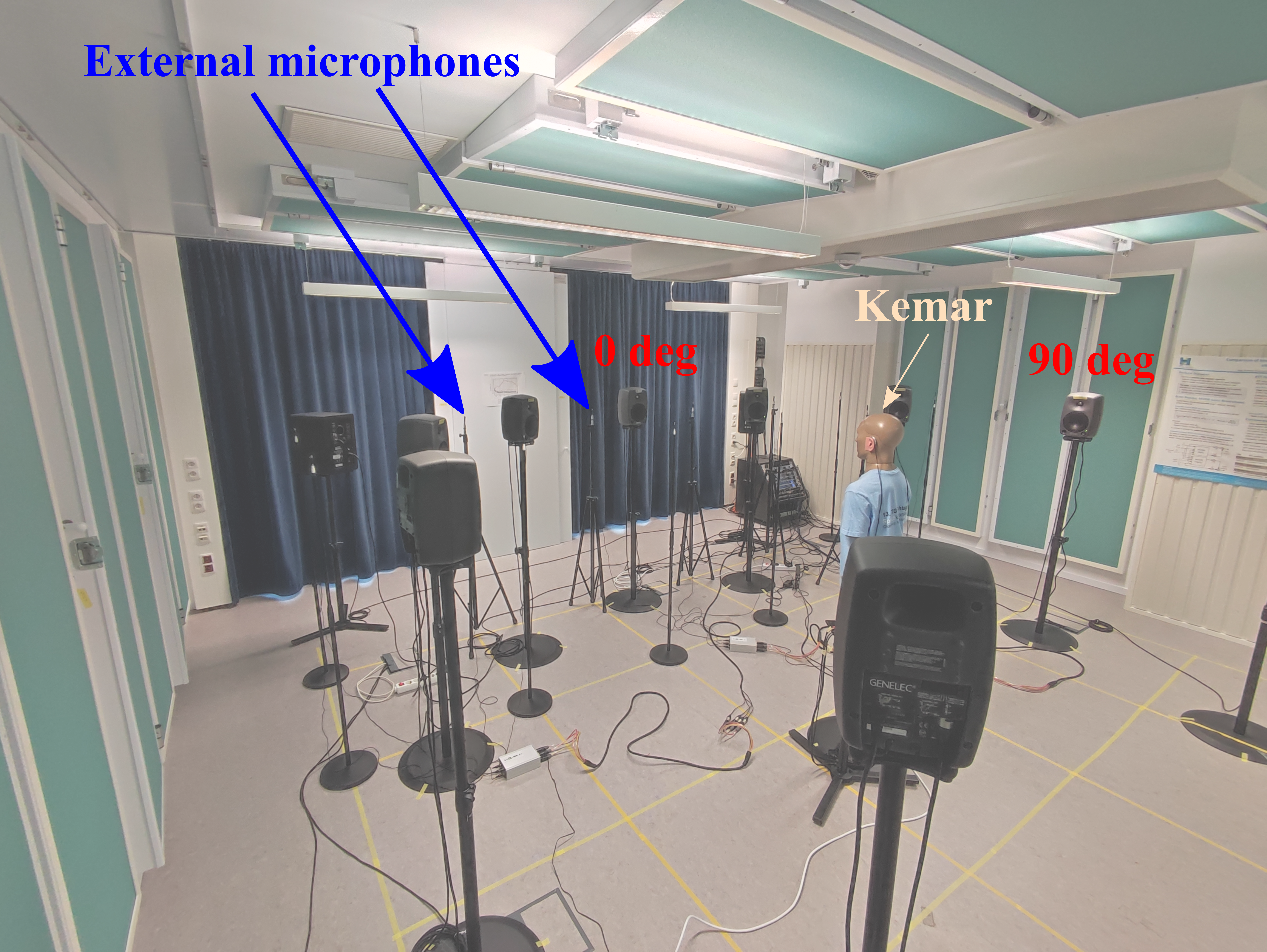}}
    \caption{\textit{Variable Acoustics Laboratory} in the low reverberation condition, with all absorber panels open and curtains closed. The picture  shows external microphones located on the positions E25-E36 (see Figure \ref{fig:Setup}).}
    \label{fig:Photo}
\end{figure}

\begin{figure}[t]
    \centering
    \trimbox{0.25cm 0.1cm 0.15cm 0cm}{\input{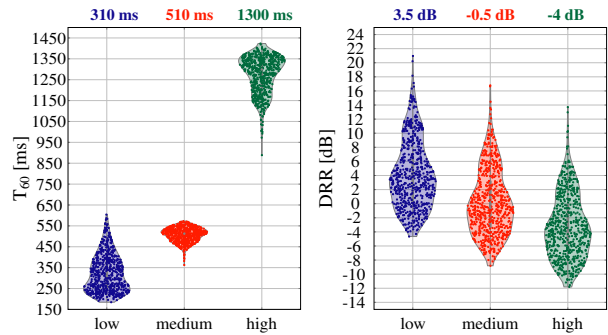}}%
    \vskip-0.15cm
    \caption{Violin plots of the distribution of the reverberation time $\mathrm{T}_{\mathrm{60}}$ (left) and the direct-to-reverberation ratio (right) for the three considered reverberation conditions. Median values are shown on top of the figure.}
    \label{fig:histo_T60_DRR}
    \vskip-0.2cm
\end{figure}

All measurements were performed in the \textit{Variable Acoustics La\-boratory} at the University of Oldenburg that has dimensions of about \qtyproduct[product-units = bracket-power]{7 x 6 x 2.7}{\metre^{3} (see~~Figure~~\ref{fig:Photo})}. The measure\-ments were per\-formed for three reverberation conditions referred to as low, medium and high. To quantify the amount of reverberation, we consider the reverberation time $\mathrm{T}_{\mathrm{60}}$ as well as the direct-to-reverberation ratio (DRR). The reverberation time of each RIR was determined by extrapolating the decay time from $\qty{-5}{\decibel}$ to $\qty{-35}{\decibel}$ on the logarithmic energy decay curve, obtained using Schroeder’s backward integration method \cite{Schroeder1965}. The DRR of each RIR was obtained as the ratio of the power of the direct-path component and the power of the early and late reflections \cite{Naylor2010}. For each reverberation condition, Figure \ref{fig:histo_T60_DRR} depicts violin plots of the distribution of $\mathrm{T}_{\mathrm{60}}$ and DRR, obtained from the measured RIRs of all loudspeaker-microphone pairs (see Sections \ref{subsec:recording__geometry} and 
\ref{sec:recording__signalAcquisition}). For each reverberation condition, the distributions of $\mathrm{T}_{\mathrm{60}}$ and DRR clearly show strong variations for different loudspeaker-microphone pairs, mainly depending on their distance and their positions relative to the walls. The median values for these reverberation parameters are $\Tsix\approx \{310, 510, 1300\}$ \unit{\milli\second} and ${\rm DRR} \approx \{3.5, -0.5, -4.0\}$ \unit{\decibel}.

\subsection{Microphone and Loudspeaker Configuration}\label{subsec:recording__geometry}
As mentioned above, Figure \ref{fig:Setup} depicts the positions of the microphones, the loudspeakers, and the dummy head. We considered head-mounted microphones as well as spatially distributed microphones, which we refer to as eMics. The six head-mounted microphones consist of four microphones of a binaural hearing aid (two microphones on each hearing aid with an an inter-microphone distance of about \qty{15}{\milli\meter}) mounted on a G.R.A.S. KEMAR 45 BM dummy head with anthropometric pinnae and two low noise in-ear microphones. It should be noted that the used hearing aids are the same as the hearing aids in the database \cite{Kayser2009}, where we considered the front and rear microphones. The dummy head was located approximately in the center of the laboratory, with its ears at an approximate height of \qty{1.5}{\meter}.

Besides the head-mounted microphones, a total of 36 eMic positions were considered in this database. Omnidirectional microphones (type: Sennheiser KE 4-211-2) were placed on a uniform \qtyproduct[product-units = bracket-power]{5 x 5}{\metre^{2}} grid around the dummy head with an inter-microphone spacing of $\qty{95(2)}{\centi\meter}$ in x- and y-direction at a height of about \qty{1.5}{\meter}. The positions of the eMics were manually calibrated by means of a cross-line laser and a laser distance meter, resulting in a positioning error of about \qty{2}{\centi\meter}.

For the measurement of the RIRs and the recording of the speech signals, 12 loudspeakers (types GE\-NE\-LEC 8030A, 8030B and 8330 APM) were placed at a distance of about $\qty{200(2)}{\centi\meter}$ at angles $\qty{-180(2.5)}{\degree}$, $\qty{-150(2.5)}{\degree}$,$\dotsc$,$\qty{150(2.5)}{\degree}$, relative to the dummy head. Each loudspeaker was set to a height of approximately $\qty{1.5}{\meter}$. The distance between the loudspeakers and the dummy head was verified using a laser distance meter. Based on three procedures with a resolution of \qty{5}{\degree}, the angles between the loudspeakers relative to the dummy head were verified \cite{Schmidt19886,Kayser2014,Fejgin2021emic}.

For the recording of the noise signals, four loudspeakers (M-Audio BX8 D2) were placed at about \qty{170}{\centi\meter} from (and facing) the corners of the room.

\section{Measurement Procedure}\label{sec:recording__signalAcquisition}
All measurements were performed at a sampling rate of \qty{48}{\kilo\hertz} using an RME MADIface XT audio interface and RME ADI-8 QS and RME OctaMic XTC analog-digital-converters. For all recordings, the input-output delay due to the buffering of the audio interface was compensated. For the RIR measurements, we used the ex\-ponential sine sweep technique \cite{Farina2000} using synchronized swept sine signals \cite{Novak2010} as excitation signals. The excitation signal had a duration of about \qty{5}{\second} and the frequency increased exponentially in the range \qty{20}{\hertz} -- \qty{24}{\kilo\hertz}. Playing back the excitation signal with each of the loudspeakers separately, the excitation signal was recorded 10 times with all available microphones simultaneously and averaged. By convolving this average recording with the inverse sweep and considering only the linear part of the resulting set of higher-order impulse responses, the RIRs between the loudspeaker and the microphones were obtained. To smoothen and truncate the RIRs to a duration of \qty{2}{\second}, Hanning windows with a duration of \qty{0.38}{\milli\second} and \qty{50}{\milli\second} for the fade-in and the fade-out, respectively, were applied.

As only 12 identical microphones were available for the spatially distributed microphones, for each reverberation condition we performed three separate recordings to cover all eMic positions E1 - E36. We refer to each separate recording as a run. We only changed the position of these 12 eMics after performing all measurements for all reverberation conditions (since changing the reverberation condition is more reproducible than changing the microphone position). In each run we recorded 18 channels simul\-taneously, i.e., the six head-mounted microphones (at the same position for all runs) and 12 eMics (run 1: positions E1 - E12, run 2: positions E13 - E24, and run 3: positions E25 - E36). Since due to scattering effects the positioning of the eMics may affect the signals recorded at the head-mounted microphones, the database provides two types of RIRs for the head-mounted microphones, which we refer to as individual and average. For the individual head-mounted RIRs the measured responses were considered for each run individually. For the average head-mounted RIRs the measured recordings were averaged over the three runs, (i.e. run-averaged) before performing the convolution with the inverse sweep.

The multi-channel speech recordings correspond to two female and two male English speech signals (duration: 20 or \qty{30}{\second}) from the \mbox{SQAM} and \mbox{VCTK} corpus \cite{ebuSQAM,veaux2017VCTK}. Similarly as for the RIRs, each speech signal was played back with each of the loudspeakers separately, and recorded with all available microphones simultaneously. The multi-channel noise recordings correspond to babble noise, cafeteria noise, and white noise (duration: \qty{20}{\second}). The noise was generated with the four loudspeakers facing the corners of the laboratory, playing back different (uncorrelated) versions of the respective noise. Similarly as for the RIRs, for the speech and noise recordings we also distinguish the individual and run-averaged recordings for the head-mounted microphones.

\section{Database}
\label{sec:availability}
The BRUDEX database is available under \url{https://doi.org/10.5281/zenodo.7986447} under a Creative Commons Attribution Non Commercial 4.0 International License. The measured RIRs and the speech and noise re\-cordings are located in sub-directories organized by the reverberation condition. The file names encode the DOA of the used loudspeaker (not applicable for noise recordings) and indicate whether the head-mounted microphone signals are obtained as individual recordings (18 channel signals) or as run-averaged recordings (42 channel signals). For the former case, the data contains 18 channels corresponding to the six head-mounted microphones and the 12 eMics. For the latter case, the data contains 42 channels corresponding to the six head-mounted microphones and the eMics at the 36 positions. The content of the database is organized in separate uncompressed binary MAT-files. To facilitate usage of the database, we provide MATLAB and Python scripts for accessing the multi-channel data.

\section{Application to MVDR Beamforming}
\label{sec:application}
To demonstrate the usability of the BRUDEX database, in this section we consider an exemplary multi-microphone noise reduction application. More particularly, we consider a binaural minimum variance distortionless response (MVDR) beamformer using the hearing aid microphones and three external microphones. Besides using estimated relative transfer function vectors to steer the MVDR beamformer, we furthermore investigate the performance of different database RIRs to investigate the robustness against non-matching RIRs. We briefly review the used algorithms in Section \ref{subsec:algos}, the considered acoustic scenario and the implementation in Section \ref{subsec:scene}, and the simulation results in Section \ref{subsec:results}.

\subsection{Algorithms}\label{subsec:algos}
We consider an acoustic scenario with a single target speaker and quasi-diffuse background noise and $M$ microphones composed of $M_\hsub=4$ hearing aid microphones and $M_\esub=3$ external microphones.
In the short-time Fourier transform (STFT) domain, with $k$ denoting the frequency bin index and $l$ denoting the frame index, the $M$-dimensional noisy signal vector $\y(k,l)$ can be written as
\begin{equation}\label{eq:yvec}
    \y(k,l) = \x(k,l) +\n(k,l) = \h(k,l) X_\refsub(k,l) + \n(k,l)\,,
\end{equation}
where $\x(k,l)$ and $\n(k,l)$ denote the speech and noise component, respectively. 
We assume a multiplicative transfer function model \cite{Avargel2007multiplicative} such that the speech component can be written in terms of its relative transfer function (RTF) vector $\h(k,l)$, which relates the speech component in a reference microphone  $X_\refsub(k,l)$ to the speech components in all other microphones. In the following, the indices $k$ and $l$ are omitted wherever possible.
By assuming un\-correlated speech and noise components, the noisy covariance matrix $\Ry = \mathcal{E}\{\y\y^H\}$ can be decomposed into the speech co\-variance matrix $\Rx$ and the noise covariance matrix $\Rn$, i.e.,
\begin{equation}\label{eq:covmat}
    \Ry = \Rx +\Rn \, ,
\end{equation}
where $\mathcal{E}\{\cdot\}$ denotes the expectation operator and $\{\cdot\}^H$ denotes the Hermitian transpose operator.
The MVDR beamformer \cite{Doclo2015,Gannot2017,VanVeen1988a} aims at minimizing the noise power spectral density (PSD) while preserving the speech component in the reference microphone. It was shown in \cite{Doclo2015,Gannot2017} that the MVDR beamformer is given by
\begin{equation}\label{eq:MVDR}
    \w = \frac{\Rnest\inv\hest}{\hest^H\Rnest\inv\hest}\, ,
\end{equation}
which requires estimates of the RTF vector $\hest$ and the noise co\-variance matrix $\Rnest$.

To estimate the RTF vector for the MVDR beamformer in \eqref{eq:MVDR}, in \cite{Goessling2018_iwaenc_a} and \cite{GoesslingWASPAA2019} methods were proposed that exploit one or mul\-tiple external microphones, assuming that the noise component in each external microphone signal is uncorrelated with the noise components in all other microphone signals. This assumption holds quie well, e.g., for a diffuse noise field where the external microphones are spatially separated from each other and from the hearing aid microphones \cite{Goessling2018_iwaenc_a}.  The so-called spatial coherence (SC) method proposed in \cite{Goessling2018_iwaenc_a} estimates the RTF vector using the $m_\esub$-th eMic by simply selecting the $M_{\mathrm{H}}\! +\! m_\esub$-th column of the noisy covariance matrix and dividing it by the reference entry, i.e.,
\begin{equation}\label{eq:SC1}
    \hest^{\SC}_{ m_\esub} = \frac{\Ryest\evec_{m_\esub}}{\evec_\refsub^T\Ryest\evec_{m_\esub}}
\end{equation}
where $\evec_{m_\esub}$ denotes a selection vector for the $m_\esub$-th eMic and $\evec_\refsub$ denotes a selection vector for the reference microphone(s), i.e., one for the left and right hearing aid each to allow for binaural processing.
Since for each of the $M_\esub$ available eMics a different RTF vector estimate $\hest^\SC_{m_\esub}$ is obtained, it was proposed in \cite{GoesslingWASPAA2019} to linearly combine them as $\hest = \Hest\avec$, where $\Hest$ is a matrix containing all estimates, i.e., 
\begin{equation}\label{eq:Hmat}
    \Hest = [\hest^{\SC}_1,\dots,\hest^{\SC}_{ M_\esub}]\,.
\end{equation} 
with $\avec$ denoting a weight vector.
A particular linear combination proposed in \cite{GoesslingWASPAA2019} aims at maximizing the output SNR of the MVDR beamformer by means of the weight vector $\avec^{\mSNR}$. The weight vector of the so-called mSNR method is given by \cite{GoesslingWASPAA2019}

\begin{equation}\label{eq:OptProb}
    \avec^{\mSNR} =\frac{\mathcal{P}\{\Bmat\inv\Amat\}}{\ones{M_\esub}{1}^T\mathcal{P}\{\Bmat\inv\Amat\}}\,,
\end{equation}
where $\mathcal{P}\{\cdot\}$ denotes the principal eigenvector operator, $\ones{M_\esub}{1}$ denotes an $M_\esub$-dimensional vector of ones and $\Bmat = \Hest^H\Rnest\inv\Hest$ and $\Amat = \Hest^H\Rnest\inv\Ryest\Rnest\inv\Hest$. The RTF vector estimate using the mSNR method is obtained as $\hest^\mSNR = \Hest\avec^\mSNR$. For further details, we refer the reader to the literature mentioned above.

\subsection{Acoustic Scenario and Implementation}\label{subsec:scene}

To evaluate the performance of different MVDR beamformers, we consider an acoustic scenario with a female target speaker (SQAM) at \qty{-60}{\degree} relative to the dummy head and quasi-diffuse babble noise in the medium reverberation condition. In addition to the four hearing aid microphones (H1-H4), we considered three eMics at the positions E14, E27 and E28. For the hearing aid microphones, we used the run-averaged RIRs and recordings.

All microphone signals were generated by using the recorded speech and noise components from the database at a broadband input signal-to-noise ratio (SNR) of \qty{0}{\decibel} in the front microphone on the left hearing aid by scaling the noise components accordingly. The resulting input SNR in the front microphone on the right hearing aid was \qty{-1}{\decibel} and in the eMics about \qty{0}{\decibel} to \qty{1}{\decibel}.

The processing was implemented in the STFT domain using a square-root-Hann window for analysis and synthesis with a frame length of 2048 samples (corresponding to about \qty{42}{\milli\second} at a samp\-ling rate of \qty{48}{\kilo\hertz}) and \qty{50}{\percent} overlap. The covariance matrices were estimated using recursive smoothing, i.e.,
\begin{equation}\label{eq:Ryest}
    \Ryest(k,l) = 
      \alpha_y\Ryest(k,l-1) + (1-\alpha_y)\y(k,l)\y^H(k,l)\, ,
\end{equation}
\begin{equation}\label{eq:Rnest}
    \Rnest(k,l) = 
      \alpha_n\Rnest(k,l-1) + (1-\alpha_n)\y(k,l)\y^H(k,l)\, ,
\end{equation}
where the parameters $\alpha_y$ and $\alpha_n$ correspond to smoothing times of \qty{500}{\milli\second} and \qty{1000}{\milli\second}, respectively.
The noisy covariance matrix $\Ryest$ was estimated during speech activity, i.e., if ${\rm \overline{SPP}}(k,l)>0.5$, whereas the noise covariance matrix $\Rnest$ was estimated during speech pauses, i.e., if ${\rm \overline{SPP}}(k,l)<0.5$. ${\rm \overline{SPP}}$ denotes the average estimated speech presence probability \cite{Gerkmann2012} in the eMics.

In the evaluation, we compare the individual RTF vector estimates obtained from the SC method in \eqref{eq:SC1} (denoted as SC-E14, SC-E27 and SC-E28, respectively) and the mSNR estimate obtained using the weight vector in \eqref{eq:OptProb}. 
To investigate the robustness against non-matching RTFs, we additionally consider the RTF vector obtained from the following database RIRs:
\begin{itemize}
    \item RIR (medium): RIRs for all used microphones for $\mathrm{DOA}=\qty{-60}{\degree}$ in the medium reverberation condition, i.e., matching the generated microphone signals.
    \item RIR (low): RIRs for all used microphones for $\mathrm{DOA}=\qty{-60}{\degree}$ in the low reverberation condition, i.e., not matching the ge\-ne\-rated microphone signals.
    \item RIR (anechoic): anechoic RIRs from \cite{Kayser2009} for the hearing aid microphones for $\mathrm{DOA}=\qty{-60}{\degree}$, i.e., an approximation with no reverberation. It should be noted that in this case only the RIRs of the hearing aid microphones and not for the eMics are available, such that the MVDR beamforming is only using the hearing aid microphones.
\end{itemize}
The RTFs were computed from the RIRs by convolving the RIRs with white Gaussian noise and computing the principal eigenvector of the resulting covariance matrix in the STFT domain.

To assess the performance of the different beamformers, we consider the broadband binaural SNR improvement ($\Delta \rm BSNR$) of the binaural MVDR beamformer compared to the unprocessed noisy reference microphone signals. 
The SNRs are computed in overlapping segments of \qty{1}{\second} length  (overlapping by 85\%) during speech activity and averaged over time. To allow for a suitable initialization of the covariance matrices, the first \qty{1.5}{\second} of the signal are not taken into account in the computation of the SNR.

\subsection{Simulation Results}\label{subsec:results}

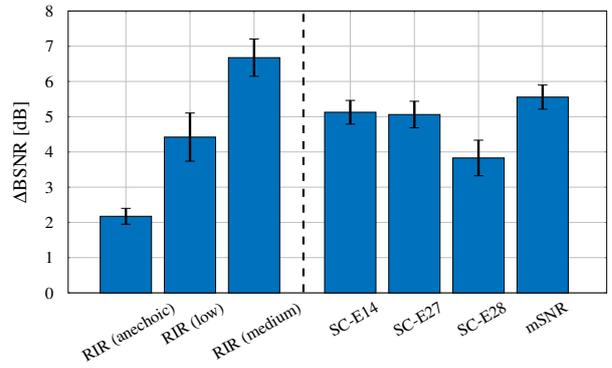
\begin{figure}[t]
    \trimbox{0.25cm 0cm 0cm 0cm}{
%
%
\definecolor{mycolor1}{rgb}{0.00000,0.44700,0.74100}%
\begin{tikzpicture}[scale = 0.18]

\begin{axis}[%
width=15.5in,
height=8.175in,
at={(2.6in,1.103in)},
scale only axis,
bar shift auto,
xmin=0.1,
xmax=8.5,
xtick={{1},{2},{3},{4.5},{5.5},{6.5},{7.5}},
xticklabels={{RIR (anechoic)},{RIR (low)},{RIR (medium)},{SC-E14},{SC-E27},{SC-E28},{mSNR}},
xticklabel style={rotate=30},
xticklabel style={font=\color{black}, font = \fontsize{35}{1}\selectfont,align=center,yshift=-0.5cm},
xlabel style={font=\color{black}, font = \fontsize{40}{1}\selectfont},
ymin=0,
ymax=8,
ytick={{0},{1},{2},{3},{4},{5},{6},{7},{8}},
yticklabels={{0},{1},{2},{3},{4},{5},{6},{7},{8}},
ylabel={$\Delta\text{BSNR [dB]}$},
ylabel style={font=\color{black}, font = \fontsize{40}{1}\selectfont,yshift=-2.5cm},
yticklabel style={font=\color{black}, font = \fontsize{35}{1}\selectfont,align=center,text width=7.5cm,xshift=2.5cm},
axis background/.style={fill=white},
xmajorgrids,
ymajorgrids,
]
\addplot[ybar, bar width=0.8, fill=mycolor1, draw=black, area legend] table[row sep=crcr] {%
1	2.17343621440506\\
2	4.4209439228339\\
3	6.67584552035537\\
4.5	5.12640111771967\\
5.5	5.06166144862177\\
6.5	3.82884319014307\\
7.5	5.55592687559049\\
};

\addplot [color=black, line width=1.0pt, draw=none,error bars/.cd, y dir=both, y explicit,
error bar style={line width=5pt,solid},error mark options={line width=1pt,mark size=10pt,rotate=90}]
 table[row sep=crcr, y error plus index=2, y error minus index=3]{%
1	2.17343621440506	0.225588399452568	0.225588399452568\\
2	4.4209439228339	0.683194212444482	0.683194212444482\\
3	6.67584552035537	0.529023815924332	0.529023815924332\\
4.5	5.12640111771967	0.336059317791394	0.336059317791394\\
5.5	5.06166144862177	0.37648072836691	0.37648072836691\\
6.5	3.82884319014307	0.50513437056919	0.50513437056919\\
7.5	5.55592687559049	0.34197184314853	0.34197184314853\\
};

\end{axis}

\draw [thick,dashed] (23.8,2.85) -- (23.8,23.6);

\end{tikzpicture}%
}%
    \vskip-0.15cm
	\caption{Improvement of the BSNR for the RTF-vector-steered binaural MVDR beamformer using different RTF vector estimation methods. Left: computed from database RIRs, right: estimated from microphone signals.}
	\label{fig:dSNR}
    \vskip-0.2cm%
\end{figure} 

Figure \ref{fig:dSNR} depicts the BSNR improvement (along with the variance as error bars) for the different considered RTF vectors.

For the RTF vectors computed from database RIRs, i.e., RIR (anechoic), RIR (low) and RIR (medium), the following obser\-vations can be made: Using the RIRs from the matching reverberation condition yields the best performance (about \qty{7}{\decibel} improvement in BSNR), while using the RIRs from the low reverberation condition still yields a rather good performance (about \qty{4}{\decibel} improvement in BSNR). Yet, the performance decreases compared to the true RIRs (RIR (medium)), as RIR (low) does not account for all reflections. 
Using the anechoic RIRs yields a rather low BSNR improvement of only \qty{2}{\decibel}. This can be explained by the fact that on the one hand only the head-mounted microphones are used in the beamformer and on the other hand by the limitation of the anechoic RIRs to only approximate the direct path but no reflections.

Using the RTF vectors estimated with the SC method (SC-E14, SC-E27 and SC-E28), results in a BSNR improvement of about \qtyrange[range-units = single,range-phrase = --]{4}{5}{\decibel}. The mSNR approach yields a BSNR im\-prove\-ment of about \qty{5.5}{\decibel}, slightly but consistently outperforming the \mbox{single} SC-based RTF vector estimates, but yielding a lower BSNR improvement than using the true RIRs (RIR (medium)). These results correspond well with prior results from literature \cite{GoesslingICASSP2019,GoesslingWASPAA2019}, which indicates a good accordance of the BRUDEX database with established simulation paradigms.

\section{Summary}
\label{sec:conclusion}
In this paper, we presented the BRUDEX data\-base, a novel database of RIRs and speech and noise recordings for binaural hearing aids and uniformly spaced distributed microphones. We presented the used measurement and calibration procedures and characterized the resulting RIRs in terms of $\mathrm{T}_{\mathrm{60}}$ and $\mathrm{DRR}$. 
The applicability of the database was demonstrated by means of an MVDR beamformer where the underlying acoustic scenario was generated from the database. The RTF vectors for steering the beamformer were either estimated using two recently proposed methods that exploit external microphones or using RIRs from the database.
The database can be used for many other signal processing tasks such as source localization, dereverberation, and noise reduction.

\small
\bibliographystyle{ieeetr}
\bibliography{myBib}


\end{document}